\documentclass[11pt]{article}

\usepackage{amsmath}
\usepackage{amssymb}
\usepackage{cite}
\usepackage{graphicx}
\usepackage[dvips]{color}
\usepackage{comment}
\usepackage{lscape}
\usepackage{caption}
\usepackage{lineno}
\usepackage{multirow}
\usepackage{amsbsy}

\begin{document}

\title{\hfill ~\\[-30mm]
       \hfill\mbox{\small }\\[30mm]
       \textbf{Confronting Four Zero Neutrino Yukawa Textures with $N_2^{}$-dominated Leptogenesis}
       }

\author{\\ Jue Zhang\footnote{E-mail: {\tt zhangjue@ihep.ac.cn}}\\ \\
  \emph{\small{}Institute of High Energy Physics, Chinese Academy of Sciences, Beijing 100049, China}}

\maketitle

\begin{abstract}
\noindent {\em} We consider a restricted Type-I seesaw scenario with four texture zeros in the neutrino Yukawa matrix, in the \emph{weak basis} where both the charged-lepton Yukawa matrix and the Majorana mass matrix for right-handed neutrinos are diagonal and real. Inspired by grand unified theories, we further require the neutrino Yukawa matrix to exhibit a similar hierarchical pattern to that in the up-type quark Yukawa matrix. With such a hierarchy requirement, we find that leptogenesis, which would operate in a $N_2^{}$-dominated scenario with the asymmetry generated by the next-to-lightest right-handed neutrino $N_2^{}$, can greatly reduce the number of allowed textures, and disfavors the scenario that three light neutrinos are quasi-degenerate. Such a quasi-degenerate scenario of light neutrinos may soon be tested in upcoming neutrino experiments.

\end{abstract} 

\thispagestyle{empty}
\vfill
\newpage
\setcounter{page}{1}

\section{Introduction}

Thanks to enormous experimental efforts in the past two decades, our knowledge about neutrinos has been greatly improved; neutrinos are light but massive, and three lepton flavors are significantly mixed \cite{PDG}. Theoretically, to account for such tiny but non-zero neutrino masses, an appealing mechanism, the so-called Type-I seesaw mechanism, was long proposed \cite{seesaw, lseesaw}. In its canonical form, three heavy right-handed (RH) neutrinos are added to the matter fields of the Standard Model (SM), and the leptonic part of Yukawa interactions is extended to

\begin{eqnarray}
-\mathcal{L}_{\text{SS}}^{} = \overline{\ell_L^{}} Y_e E H + \overline{\ell_L^{}} Y_\nu N \tilde{H}+\frac{1}{2}\overline{N^c_{}} M_R^{} N + h.c.,
\end{eqnarray}
where $\ell_L^{}$, $E$, $N$ and $\tilde{H}\equiv i\sigma_2^{} H^*$ stand for lepton doublet, charged-lepton singlet, RH	neutrino and Higgs field, respectively. The Yukawa couplings for charged leptons and neutrinos are respectively given by $Y_e^{}$ and $Y_\nu^{}$, while $M_R^{}$ denotes the Majorana mass matrix for RH neutrinos. After the Higgs field picks up its electroweak vacuum expectation value $\langle H \rangle =v \approx 174~\text{GeV}$, the charged leptons and neutrinos obtain their Dirac mass terms via $M_e^{}=v Y_e^{}$ and $M_D^{} = v Y_\nu^{}$. Due to the presence of the Majorana mass term $M_R^{}$, the left- and right-handed neutrinos are then mixed, and the mass matrix of light neutrinos is given by the famous seesaw formula $M_\nu^{} \approx M_D^{} M_R^{-1} M_D^T$, after integrating out the heavy RH neutrino fields. Such an effective neutrino mass matrix $M_\nu^{}$ then participates in the low energy neutrino phenomenology, with its eigenvalues identified as the light neutrino masses, and its mismatch with $M_e$ being the lepton mixing matrix. 

Such a Type-I seesaw mechanism is often thought to be attractive in several aspects. First, including heavy RH neutrino fields is consistent with grand unified theories (GUTs). For example, in $SO(10)$ GUTs, such RH neutrino fields, together with the SM matter fields, fit perfectly into a single $\mathbf{16}$ spinor representation of $SO(10)$. Second, given the fact that the neutrino Dirac mass term $M_D^{}$ is naturally around the electroweak scale, the mass scale of RH neutrinos is then found to be around $10^{14}~\text{GeV}$, so as to explain the sub-eV light neutrino masses. Such a mass scale of RH neutrinos is quite close to that of GUTs, $10^{15\sim16}~\text{GeV}$, and therefore the consistency with GUTs is also justified quantitatively. Lastly, it was also found that one may even employ these heavy RH neutrinos to generate the observed baryon asymmetry in the Universe (BAU) via their out-of-equilibrium and CP violating decays, a mechanism dubbed as ``leptogenesis''  \cite{lepto}.  

Given the above salient features, one then intends to probe this Type-I seesaw mechanism closely, especially with the increasingly accumulated data from low energy neutrino experiments. However, such a probe is often hindered by the large amount of free parameters possessed by the above three mass matrices $M_e^{}$, $M_\nu^{}$ and $M_R^{}$. In the weak basis (WB) where both $M_e^{}$ and $M_R^{}$ are diagonal and real, we have eighteen physical parameters in $M_D^{}$ and $M_R^{}$, however, the independent observables in low energy neutrino experiments are only nine. 
As a result, even with one more observable from leptogenesis, there always exists some parameter space in $M_D^{}$ and $M_R^{}$ so that they can agree with experiments.\footnote{Such a degeneracy in the seesaw parameter space is named as ``seesaw degeneracy'' in \cite{ssdegen}, and it can be more explicitly seen by using the so-called Casas-Ibarra parametrization \cite{CIpara}.} Testing or reconstructing a generic Type-I seesaw mechanism becomes a formidable task.


In the spirit of Occam's razor \cite{razorT,razorY}, one is then motivated to investigate some minimal scenarios with fewer parameters. One good example of such minimal scenarios is given in \cite{FGY}, where in the WB only \emph{two} sets of RH neutrinos are introduced and two texture zeros are further assumed in the neutrino Dirac mass matrix $M_D^{}$. Under these two assumptions, only five physical parameters are present in $M_D^{}$, and all of them can be determined by the five well measured low energy neutrino parameters, i.e., two neutrino mass squared differences and three lepton mixing angles, yielding predictions on those parameters that have not yet been well measured. Because of its predictive power and testability, such a minimal model has received lots of attention in the literature \cite{MX,GX,MSMreview,Goswami:2008rt,razorY}, and a systematic study considering the renormalization group running effects and the recent neutrino data can be found in \cite{Zhang:2015tea}.


In this paper, we consider an extension of the above two-generation minimal seesaw scenario by including a third set of RH neutrinos, i.e., working within the conventional seesaw framework. We similarly work in the WB, and impose texture zeros in $M_D^{}$ so as to reduce the number of free parameters.\footnote{Note that our texture analysis is restricted to such a choice of WB basis where $M_e^{}$ and $M_R^{}$ are diagonal. In principle, when searching for minimal scenarios of the Type-I seesaw mechanism, one should also consider other cases where $M_e^{}$ and(or) $M_R^{}$ are non-diagonal, see, e.g., Ref \cite{Fritzsch:2012rg}.} As was found in \cite{Branco:2007nb}, four is the maximal number of texture zeros in $M_D^{}$, assuming that none of the light neutrino masses vanishes. We will focus on this four texture zero scenario for $M_D^{}$ throughout this paper.


A simple parameter counting, however, reveals that the total number of physical parameters in $M_D^{}$ with four zeros is seven, two more than the number of well measured neutrino parameters at low energy. As a result, such a model is much less constrained compared to the above two-generation case, and testing its viability also becomes more difficult. To further constrain the model, one can impose additional theoretical assumptions, e.g., $\mu-\tau$ symmetry in $M_D^{}$ \cite{Adhikary:2009kz}, or require the model to satisfy extra experimental constraints from, e.g., leptogenesis and charged lepton flavor violation \cite{Branco:2007nb,Choubey:2008tb,Adhikary:2009kz,Adhikary:2010fa,Adhikary:2011pv,Liao:2013kix}, or a combination of both \cite{Adhikary:2009kz}. 

Here we also consider a further restriction by imposing GUT relations theoretically, and by requiring the constraint from leptogenesis experimentally.\footnote{See \cite{Kile:2013gla} for another example of studying seesaw mechanism with GUT relations imposed.} Since in $SO(10)$ the Yukawa coupling of up-type quarks is closely related to that of neutrinos, it is natural to assume that $Y_\nu^{}$ exhibits a similar hierarchy to that in the up-quark sector, and that its largest eigenvalue is close to one. One advantage of doing this is that both the overall scale and the mass spectrum of RH neutrinos are known, so that 
the formalism of leptogenesis can be readily applied. Especially, 
one knows exactly which kind of flavor approximation should be adopted in the so-called flavored leptogenesis \cite{Abada:2006fw,Nardi:2006fx,Abada:2006ea,Dev:2014laa}, see, e.g., Refs \cite{Davidson:2008bu,DiBari:2012fz} for reviews on this subject. Such a treatment on leptogenesis is in contrast to the  previous work \cite{Choubey:2008tb,Adhikary:2010fa,Liao:2013kix}, in which, without the requirement on the overall strength of $Y_\nu^{}$, one usually has to make a specific choice for the flavor approximation, and locates the mass scales of RH neutrinos accordingly.

Studies of leptogenesis within a GUT framework can also be found in the literature, e.g., Refs\cite{Buchmuller:1996pa,Nezri:2000pb,Buccella:2001tq,Branco:2002kt,Akhmedov:2003dg,DiBari:2008mp,DiBari:2010ux,DiBari:2013qja,DiBari:2014eya,Buccella:2012kc}. It was found that obtaining successful leptogenesis within the above $SO(10)$ GUT framework is not a simple matter \cite{Branco:2002kt}, as such GUT requirements on the neutrino Yukawa matrix would lead to very hierarchical RH neutrinos in general, and the lightest RH neutrino $N_1^{}$ would be too light to reproduce the observed baryon asymmetry \cite{Davidson:2002qv}. One way out is to rely on flavor effects, and to employ the next-to-lightest RH neutrino $N_2^{}$ to generate the asymmetry, with the caution that the generated asymmetry by $N_2^{}$ is not washed out by $N_1^{}$. Such a $N_2^{}$-dominated scenario has also been long discussed, e.g., in Refs \cite{DiBari:2005st,Vives:2005ra,Engelhard:2006yg}.

In this paper we adopt this $N_2^{}$-dominated leptogenesis scenario, and interestingly notice that it is very sensitive to the textures of $Y_\nu^{}$. An analysis, even at a qualitative level, is able to suggest that about half of the textures that are allowed by low energy neutrino data are unpromising to reproduce the observed baryon asymmetry. A later quantitative study further reduces such a number by a factor of two, resulting in about 15 viable patterns in the end. Moreover, due to the hierarchy and leptogenesis requirements, this scenario with four texture zeros in $Y_\nu^{}$ becomes more constrained, and disfavors the case that three light neutrinos are quasi-degenerate. Such a scenario with quasi-degenerate light neutrinos may soon be tested by upcoming neutrino experiments. 

The paper is organized as follows. In Section 2, we categorize the patterns of $Y_\nu^{}$ with four texture zeros according to their  low energy neutrino phenomenology. A further restriction of the model by GUT requirements is introduced in Section 3. In Section 4, we confront textures in such a GUT-inspired scenario with $N_2^{}$-dominated leptogenesis, where a brief review on $N_2^{}$-dominated leptogenesis is first performed, followed by a qualitative analysis and a detailed numerical study. Section 5 goes to our summary and conclusion. All the finally allowed textures can be found in Appendix A.

\section{Neutrino Yukawa Matrix with Four Texture Zeros}

In the WB where $Y_e^{}$ and $M_R^{}$ are diagonal, there are 126 possible ways of assigning four texture zeros in the neutrino Yukawa matrix $Y_\nu^{}$. Barring the cases that have block diagonal forms, and that yield zero neutrino masses at low energy, the total number of remaining patterns is 72. These 72 patterns can be classified into two major categories, according to their resulting properties on the effective neutrino mass matrix $M_\nu^{}$ \cite{Branco:2002kt}. Each of these two categories further contains three subclasses. 

\begin{itemize}

\item \textbf{Category I} -- Two rows of $Y_\nu^{}$ are orthogonal, resulting in one texture zero in the off-diagonal entries of $M_\nu^{}$. Three subclasses IA, IB and IC are defined as the cases with one zero in $(M_\nu^{})_{23}$, $(M_\nu^{})_{13}$ and $(M_\nu^{})_{12}$, respectively. For Class IA, we have three distinct types of textures that keep the last two rows of $Y_\nu^{}$ orthogonal, namely, 

\begin{eqnarray} \label{eq:IA}
\textbf{Class IA:}\qquad
\begin{pmatrix}
\times & \times & \times \\
\times & \mathbf{0} &\mathbf{0} \\
\mathbf{0} & \times & \mathbf{0}
\end{pmatrix},\qquad
\begin{pmatrix}
\times & \times & \mathbf{0} \\
\times & \mathbf{0} &\times \\
\mathbf{0} & \times & \mathbf{0}
\end{pmatrix}, \qquad 
\begin{pmatrix}
\times & \times & \mathbf{0} \\
\mathbf{0} & \times & \mathbf{0}\\
\times & \mathbf{0} &\times 
\end{pmatrix},
\end{eqnarray}
where the cross `$\times$' denotes a non-vanishing entry. Due to the freedom of shuffling the order of RH neutrinos, each of the above matrix can have further six possible permutations on its three columns, resulting in 18 patterns for Class IA in total. Textures in Class IB and IC can be similarly obtained by interchanging rows of $Y_\nu^{}$ in Class IA.

\item \textbf{Category II} -- Two columns of $Y_\nu^{}$ are orthogonal and one column is without zeros. Such a structure leads to one texture zero in the off-diagonal entries of $M_\nu^{-1}$. Three subclasses IIA, IIB and IIC correspond to the cases with vanishing $(M_\nu^{-1})_{23}$, $(M_\nu^{-1})_{13}$ and $(M_\nu^{-1})_{12}$, respectively, and they are given by

\begin{eqnarray}
\textbf{Class IIA:} &\qquad &
\begin{pmatrix}
\times & \mathbf{0} & \mathbf{0}\\
\times & \times & \mathbf{0} \\
\times & \mathbf{0} & \times
\end{pmatrix}, \\
\textbf{Class IIB:} &\qquad &
\begin{pmatrix}
\times & \mathbf{0} & \times\\
\times & \mathbf{0} & \mathbf{0} \\
\times & \times & \mathbf{0}
\end{pmatrix}, \\
\textbf{Class IIC:} &\qquad &
\begin{pmatrix}
\times & \times & \mathbf{0}\\
\times & \mathbf{0} & \times \\
\times & \mathbf{0} & \mathbf{0}
\end{pmatrix}, 
\end{eqnarray}
subject to six possible permutations on three columns. 

\end{itemize}

Absorbing the overall phase present in each row of $Y_\nu^{}$ by charged leptons, the physical parameters in $Y_\nu^{}$ are found to be seven, of which five are the magnitudes of those five non-zero entries, and two are the relative phases in the rows that contain more than one non-zero entries. Since in the seesaw formula, the overall magnitude of each column of $Y_\nu^{}$ is always combined with one of the masses of RH neutrinos, it would be useful to introduce a ``rescaled'' Dirac Yukawa matrix $Y$, 

\begin{eqnarray} \label{eq:Ydef}
Y = v Y_\nu M_R^{-1/2},
\end{eqnarray}
as was done in \cite{Liao:2013kix}. The low energy effective neutrino mass matrix $M_\nu^{}$ is then given by

\begin{eqnarray}\label{eq:YMnu}
M_\nu = Y Y^T.
\end{eqnarray}
Such a rescaled Dirac Yukawa matrix $Y$ inherits the texture structure of $Y_\nu^{}$, and therefore it also contains seven independent parameters. Interestingly, $M_\nu^{}$ or $M_\nu^{-1}$ with one texture zero in off-diagonal entries possesses the same amount of independent parameters, and one can check that these two sets of seven parameters in $Y$ and $M_\nu$ can be derived mutually, up to some ambiguities of signs. Because of this, the low energy neutrino phenomenology of $Y_\nu^{}$ with four texture zeros is identical to that of $M_\nu^{}$ or $M_\nu^{-1}$ with one texture zero in its off-diagonal entries. It should be noted that such one texture zero structure of $M_\nu^{}$ or $M_\nu^{-1}$ is invariant under the renormalization group (RG) running at one loop level.\footnote{See \cite{Ohlsson:2013xva} for a recent review on the RG evolution of neutrino masses and lepton flavor mixing parameters.} 

The neutrino phenomenology of $M_\nu^{}$ or $M_\nu^{-1}$ with one texture zero has also been extensively studied in the literature \cite{Liao:2013kix,Xing:2003ic,Merle:2006du,Liu:2013oxa,Lashin:2011dn,Lashin:2009yd,Dev:2010if,Liao:2013rca}. With seven independent parameters in $M_\nu^{}$, there exist two relations among the nine low energy neutrino parameters. These nine low energy neutrino parameters are: three light neutrino masses $m_i^{}$ from the diagonalization of $M_\nu^{}$, and three mixing angles $\theta_{ij}^{}$, one Dirac phase $\delta$ and two Majorana phases $\varphi_2$ and $\varphi_3$ residing in the Maki-Nakagawa-Sakata-Pontecorvo (MNSP) neutrino matrix $U$, 

\begin{eqnarray}
U = \begin{pmatrix}
c_{12}c_{13} & c_{13} s_{12} &  s_{13}e^{-i\delta} \\
-c_{23}s_{12}-c_{12}s_{13}s_{23}e^{i\delta} & c_{12}c_{23}-s_{12}s_{13}s_{23}e^{i\delta} & c_{13}s_{23} \\
s_{12}s_{23}-c_{12}c_{23}s_{13}e^{i\delta} & -c_{12}s_{23}-c_{23}s_{12}s_{13}e^{i\delta} & c_{13}c_{23}
\end{pmatrix} \mathcal{P}
\end{eqnarray}
where $\mathcal{P}= \text{diag}(1,e^{i\varphi_2/2},e^{i\varphi_3/2})$. In the basis where $Y_e^{}$ is diagonal, this $U$ diagonalizes $M_\nu^{}$ via $M_\nu^{}= U^*\text{diag}(m_1^{},m_2^{},m_3^{}) U^\dagger$. With two constraints among these nine neutrino parameters, one then can choose to express these two Majorana phases as functions of the other seven neutrino parameters. Analytic expressions of such functions can be found in \cite{Liao:2013kix}. Furthermore, with these Majorana phases predicted, one can proceed to evaluate the effective neutrino mass $M_{ee}^{}$, 

\begin{eqnarray}
M_{ee}= |m_1 U_{e1}^2 + m_2 U_{e 2}^2 + m_3 U_{e 3}^2|,
\end{eqnarray}
which is a testable observable in the neutrinoless double beta ($0\nu\beta\beta$) experiments.

\begin{table} 
\centering
\begin{tabular*}{0.85\textwidth}{  c  c c c c }
\hline
\hline
\noalign{\smallskip}
 & \multicolumn{2}{c}{Normal Hierarchy} & \multicolumn{2}{c}{Inverted Hierarchy}  \\
\noalign{\smallskip}
\hline
& bfp $\pm 1\sigma$ & $3\sigma$ range & bfp $\pm 1\sigma$ & $3\sigma$ range\\
\hline
\noalign{\smallskip}
\noalign{\smallskip}
$\theta_{12}/^\circ$ & $33.48^{+0.78}_{-0.75}$ & $31.29 \rightarrow 35.91$ & $33.48^{+0.78}_{-0.75}$ & $31.29 \rightarrow 35.91$ \\
\noalign{\smallskip}
\noalign{\smallskip}
$\theta_{23}/^\circ$ & $42.3^{+3.0}_{-1.6}$ & $38.2 \rightarrow 53.3$ & $49.5^{+1.5}_{-2.2}$ & $38.6 \rightarrow 53.3$ \\
\noalign{\smallskip}
\noalign{\smallskip}
$\theta_{13}/^\circ$ & $8.50^{+0.20}_{-0.21}$ & $7.85 \rightarrow 9.10$ & $8.51^{+0.20}_{-0.21}$ & $7.87 \rightarrow 9.11$ \\
\noalign{\smallskip}
\noalign{\smallskip}
$\delta/^\circ$ & $306^{+39}_{-70}$ & $0 \rightarrow 360$ & $254^{+63}_{-62}$ & $0 \rightarrow 360$ \\
\noalign{\smallskip}
\noalign{\smallskip}
$\frac{\Delta m_{21}^2}{10^{-5}~\text{eV}^2}$ & $7.50^{+0.19}_{-0.17}$ & $7.02 \rightarrow 8.09$ & $7.50^{+0.19}_{-0.17}$ & $7.02 \rightarrow 8.09$ \\
\noalign{\smallskip}
\noalign{\smallskip}
$\frac{\Delta m_{3l}^2}{10^{-3}~\text{eV}^2}$ & $+2.457^{+0.047}_{-0.047}$ & $+2.317 \rightarrow +2.607$ & $-2.449^{+0.048}_{-0.047}$ & $-2.590 \rightarrow -2.307$ \\
\noalign{\smallskip}
\hline
\hline
\end{tabular*}
\caption{Latest global fit results taken from \cite{Gonzalez-Garcia:2014bfa}. Note that $\Delta m_{3l}^2 \equiv \Delta m_{31}^2 >0$ for NH and $\Delta m_{3l}^2 \equiv \Delta m_{32}^2 <0$ for IH.}
\label{tb:gfit}
\end{table} 

In addition to those two constraints among nine neutrino parameters, such one texture zero structure of $M_\nu^{}$ or $M_\nu^{-1}$ also yields an inequality among those seven independent neutrino parameters \cite{Liao:2013kix}. With the two mass squared differences and three mixing angles well measured experimentally (see Table \ref{tb:gfit} for the latest global fit results), such an inequality would weakly constrain the allowed parameter space for the lightest neutrino mass $m_1$ ($m_3$) in the case of normal hierarchy (NH) (inverted hierarchy (IH)), and the Dirac phase $\delta$.

\begin{figure}
\centering
\includegraphics[scale=0.8]{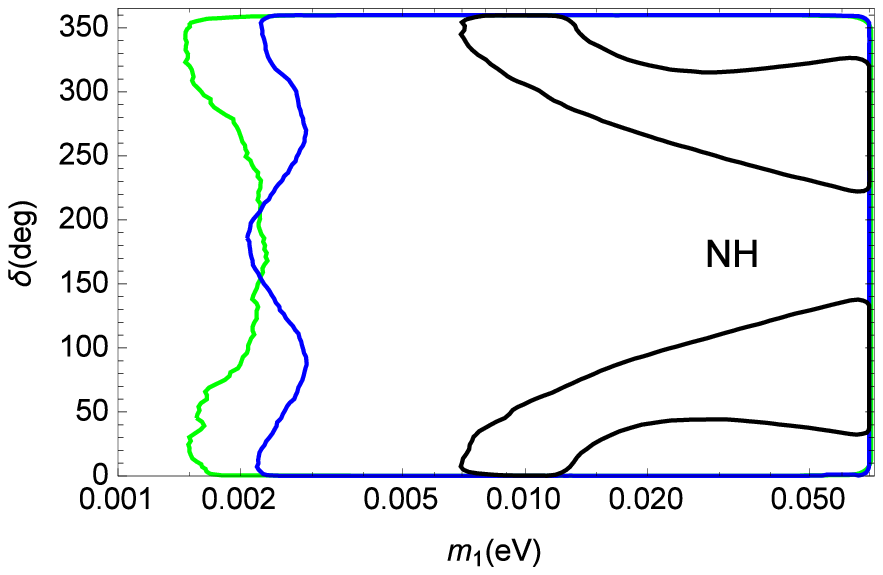}
\includegraphics[scale=0.8]{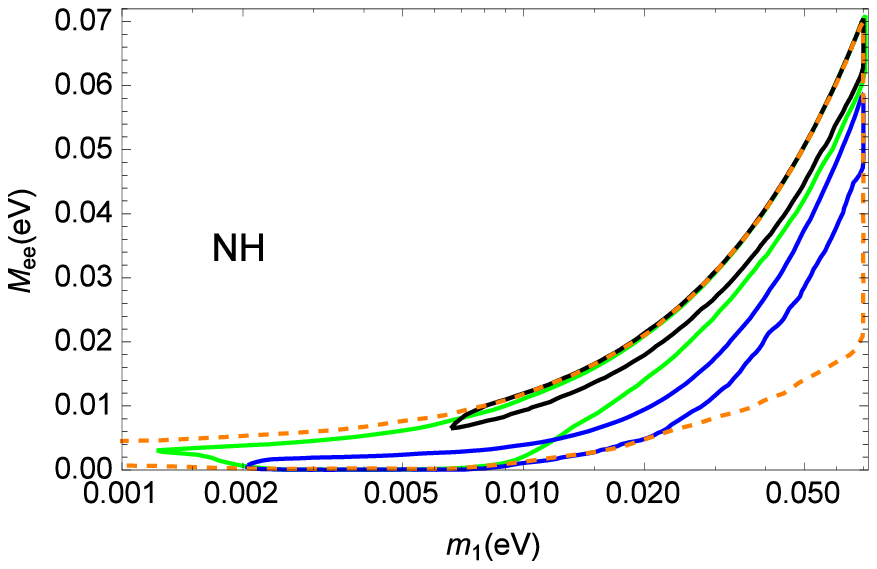}
\includegraphics[scale=0.8]{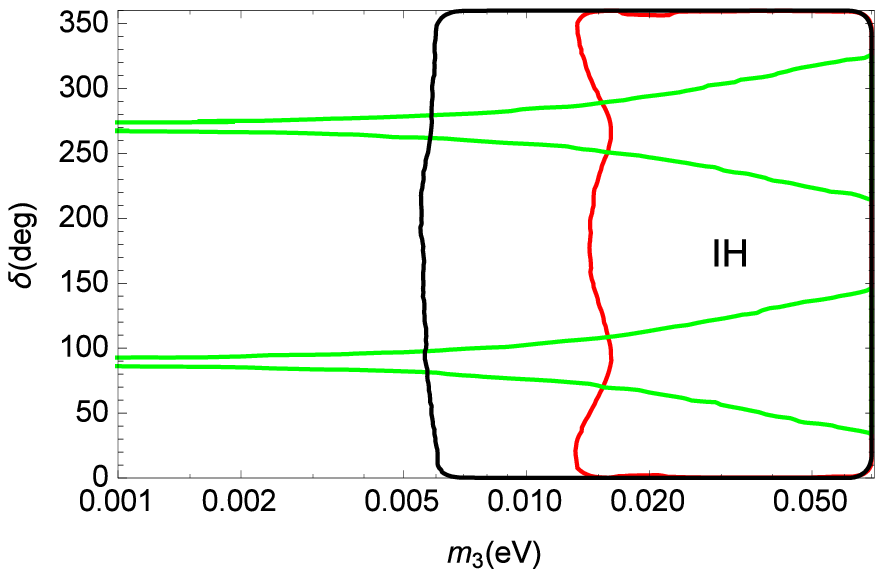}
\includegraphics[scale=0.8]{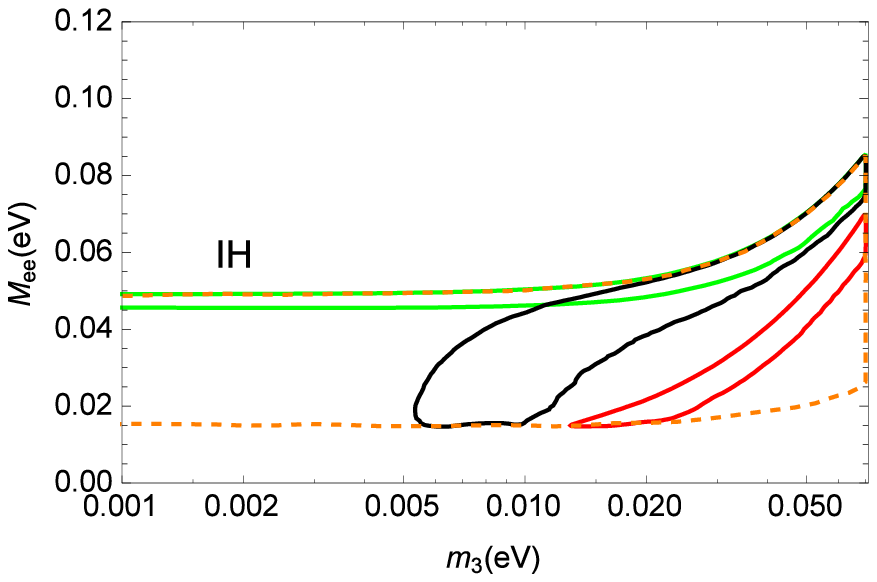}
\caption{(color online) Allowed parameter space for $m_{1,3}^{}$, $\delta$ and $M_{ee}^{}$ in Class IA (red), Class IB (green), Class IIA (blue) and Class IIB (black). Left and right panels give the parameter space in terms of ($m_{1,3}^{}$ \emph{vs} $\delta$) and ($m_{1,3}^{}$ \emph{vs} $M_{ee}^{}$), respectively, while the top (NH) and bottom (IH) panels are distinguished by two different mass hierarchies of neutrinos. Note that the orange dashed contours on the right panels indicate the allowed regions for ($m_{1,3}^{}$ \emph{vs} $M_{ee}^{}$), given the current $3\sigma$ data from neutrino oscillation experiments. }
\label{fg:LE}
\end{figure}

In Fig.\ref{fg:LE} we reproduce the numerical results given in \cite{Liao:2013kix}, by presenting the allowed parameter space for $m_{1,3}^{}$, $\delta$ and $M_{ee}^{}$ for the previously defined classes. The top and bottom panels correspond to the cases of NH and IH respectively, while the allowed parameter space for ($m_{1,3}^{}$ \emph{vs} $\delta$) and ($m_{1,3}^{}$ \emph{vs} $M_{ee}^{}$) are respectively shown on the left and right panels. In producing Fig.\ref{fg:LE}, we take the $3\sigma$ range values for the two mass squared differences and three neutrino mixing angles, and also impose the total neutrino mass bound from Planck \cite{Ade:2013zuv}, $\sum m_i < 0.23~\text{eV}$, which leads to $m_{1,3}^{} \in [0, 0.07]~\text{eV}$. Boundaries of allowed regions are represented by solid contours, and different classes are distinguished by different line colors.\footnote{In order to make a comparison of different classes, we have collected their numerical results at a single place. Then since the allowed parameter space for different classes are usually overlapped, presenting scatter plots may not be appropriate, as otherwise identifying boundaries of overlapped regions would be very challenging.} As was pointed out in \cite{Liao:2013kix}, the low energy neutrino phenomenology  of Class IC and Class IIC is similar to that of Class IB and Class IIB respectively, due to the fact that $\theta_{23} \sim 45^\circ$. We then only show the results for Class IA (red), Class IB (green), Class IIA (blue) and Class IIB (black), and the allowed parameter space for Class IC and IIC can be roughly obtained by taking $\delta \rightarrow \delta +180^\circ$. 

We next discuss the numerical results that are shown in Fig.\ref{fg:LE}, and focus on the prospect of excluding these classes regarding upcoming experimental results, including a determination of the Dirac phase $\delta$ and various methods of probing the neutrino mass scale, such as a kinematic measurement of the effective neutrino mass $m_\beta^{}$ from beta decay experiments, a measurement of the total neutrino mass from cosmology, and finally a measurement of $M_{ee}^{}$ from $0\nu\beta\beta$ experiments.\footnote{See \cite{deGouvea:2013onf} for a outlook on neutrinos physics in the next two or three decades.} The discussion will be split into two parts, with regards to the neutrino mass hierarchy, which, however, might be soon known with upcoming neutrino oscillation experiments \cite{Itow:2001ee,Ayres:2004js,Li:2013zyd,Park:2014sja,Aartsen:2014oha}.

\begin{itemize}

\item \textbf{NH:}
\begin{itemize}
\item Class IA (not shown in Fig.\ref{fg:LE}) is excluded, due to the fact that its minimal allowed value for $m_1^{}$ is around $0.15~\text{eV}$, which is in contradiction with the above cosmological bound.

\item All the other classes predict a minimal mass of $m_1^{}$ to be below $0.01~\text{eV}$, which is, however, unlikely to be excluded shortly, assuming that the pattern of negative results in various neutrino mass measurements continues. In the kinematic mass measurement, the current commissioning KATRIN experiment \cite{Osipowicz:2001sq,fortheKATRIN:2013saa} has its projected exclusion limit of only $\sim 0.2 ~\text{eV}$ on the effective mass $m_\beta^{}$, which corresponds to $\sim 0.2 ~\text{eV}$ on $m_1^{}$. In the far future, even though Project 8 \cite{Doe:2013jfe} and PTOLEMY \cite{Betts:2013uya} may push the limit of $m_\beta^{}$ down to the $\sim 0.05~\text{eV}$ level, a value equivalent to $\sim 0.05~\text{eV}$ on $m_1$, one is still unable to exclude any of these classes. As for the constraint from cosmology, a mass of $m_1^{}$ around $0.01~\text{eV}$ corresponds to a total neutrino mass of  $\sim 0.07~\text{eV}$. Although in the next decade one may push the exclusion limit of total neutrino mass down to the level of $\sim 0.05~\text{eV}$ \cite{Hannestad:2010kz,Wong:2011ip}, the obtained confidence level  might not be high enough to exclude these classes, as $m_1^{}$ is very sensitive to the value of total neutrino mass when the latter falls into the region of around $0.07~\text{eV}$. 

On the other hand, since the region with $m_1^{} \gtrsim 0.01~\text{eV}$ is allowed by all of these classes, positive signals from upcoming neutrino mass measurements are also unable to exclude any of them. 

\item Among all of the allowed classes, only Class IIB (and Class IIC) might be excluded by a measurement of $\delta$ that is close to $0$ or $180^\circ$. However, such a region of $\delta$ is quite hard to probe in the forthcoming long-baseline neutrino oscillation experiments \cite{deGouvea:2013onf}. Therefore, we do not expect that any of these classes would be shortly excluded by a measurement of $\delta$.

\item It is well-known that excluding models in NH with bounds from $0\nu\beta\beta$ experiments is quite hard. Even for the future multi-ton scale experiments, one is only able to reach $\sim 0.01~\text{eV}$ in an optimistic perspective \cite{deGouvea:2013onf,Rodejohann:2012xd}. Therefore, none of these classes can be excluded shortly by $0\nu\beta\beta$ experiments either. 

\end{itemize}

\item \textbf{IH:}

\begin{itemize}
\item Class IIA (also not shown in Fig.\ref{fg:LE}) is excluded for the same reason as the above Class IA in NH, that is, the minimal allowed value for $m_3^{}$ is not favored by Planck.

\item Excluding classes with upcoming results from neutrino mass measurements is also unlikely, regardless of positive or negative results that will be obtained experimentally.


\item Forthcoming experimental results on $\delta$ are also unlikely to exclude any of these classes. 

\item Unlike the above NH case, the allowed region for $M_{ee}^{}$ in IH may be probed by upcoming $0\nu\beta\beta$ experiments. Class IB would be excluded, if the exclusion limit of $M_{ee}^{}$ were found to be below $\sim 0.045~\text{eV}$ in the next-generation multi-100kg scale experiments \cite{Rodejohann:2012xd}. However, neither Class IIB nor Class IA can be excluded, as their allowed ranges for $M_{ee}^{}$ cover all possible values of $M_{ee}^{}$ that are allowed by neutrino oscillation data.
\end{itemize}

\end{itemize}

In summary, given the current neutrino data, Class IA in NH and Class IIA in IH have already been excluded. However, a further exclusion with various upcoming neutrino experiments is not promising, with only one exception that Class IB (and Class IC) in IH may be shortly tested by $0\nu\beta\beta$ experiments. Such low testability is attributed to the fact that seven free parameters are present in $Y_\nu^{}$, so that the parameter space for $m_{1,3}^{}$ and $\delta$ is weakly constrained. Therefore, according to Occam's razor, this four texture zero neutrino Yukawa matrix scenario cannot be viewed as a minimal case. Although the number of free parameters cannot be further reduced by taking additional texture zeros, we may restrict those seven existing parameters with some other requirements, so that the model may become more minimal and more testable. Next, we will discuss one example of such further restricted scenarios by invoking grand unified theories.

\section{Hierarchical $Y_\nu^{}$ and $M_R^{}$ Inspired by GUTs}


We start with solving for the rescaled Yukawa matrix $Y$ from $M_\nu^{}$, according to Eq.(\ref{eq:YMnu}). Due to the fact that the same amount of free parameters are found in both of them, such a solution can always be obtained. For example, for the first pattern of Class IA in Eq.(\ref{eq:IA}), we have the rescaled Yukawa matrix given by 

\begin{eqnarray}
Y=\begin{pmatrix}
a & b & c \\
d & 0 & 0 \\
0 & e & 0
\end{pmatrix}.
\end{eqnarray}
From Eq.(\ref{eq:YMnu}), one then has

\begin{eqnarray}
M_\nu= Y Y^T = \begin{pmatrix}
a^2+b^2+c^2 & ad & be\\
ad & d^2 & 0 \\
be & 0 & e^2
\end{pmatrix},
\end{eqnarray}
which yields the following solution of $a, b, c, d$ and $e$ in terms of the elements of $M_\nu^{}$,

\begin{eqnarray}
e=\pm \sqrt{(M_\nu)_{33}}, \quad d=\pm \sqrt{(M_\nu)_{22}}, \quad b= (M_\nu)_{13}/e, \quad a=(M_\nu)_{12}/d, \quad c=\pm \sqrt{(M_\nu)_{11}-a^2-b^2}. \nonumber
\end{eqnarray}
Such conversions between elements of $Y$ and those of $M_\nu^{}$ can be preformed similarly for the other classes, and they are omitted here.

With the rescaled Yukawa matrix $Y$ specified, we are now at the position of obtaining the regular neutrino Yukawa matrix $Y_\nu^{}$ with GUT requirements imposed. Such GUT requirements are inspired by the Yukawa structure of GUT $SO(10)$ models. In a minimal GUT $SO(10)$ set-up \cite{Fukuyama:2012rw}, one assigns all the matter fields including RH neutrinos to the $\mathbf{16}$ spinor representation of $SO(10)$, while two Higgs fields are found to be in the $\mathbf{10}$ and $\mathbf{126}$ representations. Due to the Yukawa interactions $(\mathbf{16}~\mathbf{16}~\mathbf{10}_H^{})$ and $(\mathbf{16}~\mathbf{16}~\mathbf{126}_H^{})$, the mass matrices of various sectors can be expressed as

\begin{eqnarray} \label{eq:SO10}
M_u^{} &=& c_{10}^{} M_{10}^{} + c_{126}^{} M_{126}^{}, \nonumber \\
M_d^{} &=& M_{10}^{} + M_{126}^{}, \nonumber\\
M_D^{} &=& c_{10}^{} M_{10}^{} - 3 c_{126}^{} M_{126}^{}, \nonumber\\
M_e^{} &=& M_{10}^{} - 3M_{126}^{}, \\
M_L^{} &=& c_L^{} M_{126}^{},\nonumber\\
M_R^{} &=& c_R^{} M_{126}^{},\nonumber
\end{eqnarray}
where, except for $M_D^{}$, $M_e^{}$ and $M_R^{}$ that have already been defined previously, $M_u^{}$, $M_d^{}$ and $M_L^{}$ denote the mass matrices of up-type quark, down-type quark and left-hand Majorana neutrinos. As one can see, all the quark and lepton mass matrices stem from two basic mass matrices, $M_{10}^{}$ and $M_{126}^{}$, and four complex coefficients $c_{10}^{}, c_{126}^{}, c_L^{}$ and $c_{R}^{}$. Because of this, mass matrices of quarks and leptons are closely related, and in this paper we are particularly interested in the connection between $M_u^{}$ and $M_D^{}$. Three kinds of requirements are then imposed on $Y_\nu$ (or equivalently on $M_D$):

\begin{itemize}

\item The largest eigenvalue of $Y_\nu$ is assumed to be one, as it is the case for the up-type quark Yukawa coupling;

\item $Y_\nu^{}$ is hierarchical, with its three eigenvalues obeying the pattern of $(\beta\lambda^8: \alpha\lambda^4: 1)$, where $\alpha$ and $\beta$ are of order one parameters, and $\lambda =0.227$ is the Cabibbo angle around the GUT scale \cite{Ross:2007az}. Such an assumption is motivated by the observation that the three eigenvalues in $M_u^{}$, identified as three up-type quark masses, are hierarchical, with a hierarchy of $(\sim \lambda^8: \sim \lambda^4: 1)$ around the GUT scale \cite{Ross:2007az,Xing:2007fb,Antusch:2013jca};

\item The above unspecified parameters $\alpha$ and $\beta$ are further taken to be $3$ and $1/3$ respectively. This is to mimic Georgi-Jarlskog mass relations \cite{Georgi:1979df} that are observed between the down-type quark and charged-lepton sectors around the GUT scale \cite{Ross:2007az,Xing:2007fb,Antusch:2013jca}, i.e., $m_\mu^{}=3m_s^{}, m_e^{}=m_d^{}/3$, where $m_d^{}, m_s^{}, m_e^{}$ and $m_\mu^{}$ are the masses of down quark, strange quark, electron and muon, respectively. These Georgi-Jarlskog mass relations between the down-type quark and charged-lepton sectors are often realized by the presence of the $\textbf{126}$ ($\textbf{45}$) dimensional Higgs field in $SO(10)$ ($SU(5)$), whose Yukawa contributions to these two sectors are differed by a factor of $-3$, see the terms containing $M_{126}^{}$ in $M_d^{}$ and $M_e^{}$ in Eq.(\ref{eq:SO10}). Since the same difference of the factor of $-3$ is also observed between the up-type quark and neutrino Yukawa matrices, one then expects similar Georgi-Jarlskog mass relations may arise for these two sectors as well. 

\end{itemize} 

Given the above assumption that $Y_\nu^{}$ is hierarchical, the Majorana mass matrix for RH neutrinos $M_R^{}$ is then likely to possess a correlated squared hierarchy, so as to offset the strong hierarchy in $Y_\nu^{}$, resulting in a mild hierarchy in $M_\nu^{}$ or $Y$. According to Eq.(\ref{eq:Ydef}), we obtain $Y_\nu^{}$ and $M_R^{}$ as,

\begin{eqnarray} \label{eq:YMR}
Y_\nu^{} &=& y_0^{} Y \begin{pmatrix}
\lambda^y & & \\
& \lambda^x & \\
& & 1
\end{pmatrix}, \\
M_R^{} &=& v^2 y_0^2 \begin{pmatrix}
\lambda^{2y} & & \\
& \lambda^{2x} & \\
& & 1
\end{pmatrix},
\end{eqnarray}
where $y_0^{}$, $x$ and $y$ are real parameters, and $x \sim 4$ and $y \sim 8$ so as to satisfy the above GUT requirements on $Y_\nu^{}$.\footnote{Here we assume that the hierarchy pattern of the eigenvalues of $Y_\nu^{}$ is mainly due to the diagonal hierarchical matrix $\text{diag}(\lambda^y, \lambda^x, 1)$, instead of some accidental cancellations among the entries of $Y$.} Due to such a hierarchical form of $Y_\nu^{}$, its three eigenvalues can be found approximately, in expansions of $\lambda$.  Requiring them to satisfy the hierarchical patten of $(\beta\lambda^8: \alpha\lambda^4: 1)$ then determines the values of $y_0^{}$, $x$ and $y$, namely,

\begin{eqnarray}
y_0^{} &\approx & 1/\sqrt{A}, \\
x &\approx & \frac{1}{2}\log _\lambda^{} \left( \alpha^2 \left \lvert \frac{A}{B} \right \rvert\right) +4, \\
y &\approx & \frac{1}{2}\log _\lambda^{} \left( \beta^2 \left \lvert\frac{A}{C}\right \rvert \right) +8,
\end{eqnarray}
where $A$, $B$ and $C$ are determined by the elements of $Y$,

\begin{eqnarray}
A &=& (Y^\dagger Y)_{33}^{}, \\
B &=& (Y^\dagger Y)_{22}^{} - \frac{|(Y^\dagger Y)_{23}^{}|^2}{(Y^\dagger Y)_{33}^{}}, \\
C &=& (Y^\dagger Y)_{11}^{}- \frac{|(Y^\dagger Y)_{13}^{}|^2}{(Y^\dagger Y)_{33}^{}}-\frac{|C^\prime |^2}{B}, \\
C^\prime &=& (Y^\dagger Y)_{12}^{} - \frac{(Y^\dagger Y)_{23}^{} (Y^\dagger Y)_{13}^{} }{(Y^\dagger Y)_{33}^{}}.
\end{eqnarray}
Since $Y$ is constructed from $M_\nu^{}$, we then expect that the non-zero elements of $Y$ are all around the same order. A more precise estimation can then be made on $x$ and $y$, namely, $x \sim \log_\lambda(\alpha)+4 \approx 3.3$ and $y \sim \log_\lambda(\beta)+8 \approx 8.7$. With the hierarchical forms of $Y_\nu^{}$ and $M_R^{}$ given above, we next study their implications on leptogenesis.

\section{Confronting Textures of $Y_\nu^{}$ with $N_2^{}$-dominated Leptogenesis}


\subsection{$N_2^{}$-dominated Leptogenesis}

Given the hierarchical pattern of the three RH neutrinos being around $(\beta^2 \lambda^{16}: \alpha^2 \lambda^8: 1)$, a sub-eV low energy neutrino mass would indicate that the masses of three RH neutrinos are around $(10^3~\text{GeV}: 10^{10}~\text{GeV} : 10^{14}~\text{GeV}$). The lightest RH neutrino $N_1^{}$ is therefore too light to generate enough asymmetry \cite{Davidson:2002qv}, and one then has to rely on the asymmetry generated by the next-to-lightest RH neutrino $N_2^{}$, provided that the generated asymmetry by $N_2^{}$ is not washed out by $N_1^{}$ eventually. Note that leptogenesis of the heaviest RH neutrino $N_3^{}$ is not considered here, due to the assumption that the reheating temperature $T_{\text{reheat}}^{}$ satisfies $M_2^{} \lesssim T_{\text{reheat}}^{} \ll M_3^{}$.

In this $N_2^{}$-dominated scenario, because its mass $M_2^{}$ falls into the region of $ 10^9~\text{GeV} \lesssim M_2^{} \lesssim 10^{12}~\text{GeV}$, we then adopt the two-flavor approximation when calculating the asymmetry generated by $N_2^{}$. In this two-flavor approximation, $\tau$ flavor is distinguished, while the other two flavors $e$ and $\mu$ remain indistinguishable, and are treated as a single flavor. The produced $(B-L)$ asymmetry at the temperature around $M_2^{}$ is then found to be \cite{DiBari:2013qja}

\begin{eqnarray}
N_{\text{B-L}}^{T\sim M_2} \approx \epsilon_{2\tau} \kappa (K_{2\tau}) + \epsilon_{2 e+\mu} \kappa (K_{2e +\mu}),
\end{eqnarray}
where $\epsilon$, $K$ and $\kappa$ are the asymmetry factors, washout factors and efficiency factors, respectively. Within the SM, the asymmetry factor $\epsilon_{2\alpha}$ is defined as

\begin{eqnarray} \label{eq:epsilon}
\epsilon_{2\alpha} &\equiv& -\frac{\Gamma(N_2 \rightarrow H l_\alpha)-\Gamma(N_2 \rightarrow \bar{H} \bar{l}_\alpha)}{\Gamma(N_2 \rightarrow H l_\alpha)+\Gamma(N_2 \rightarrow \bar{H} \bar{l}_\alpha)},
\end{eqnarray}
and is found to be \cite{Covi:1996wh}
\begin{eqnarray} 
\epsilon_{2\alpha} &\approx & \frac{1}{8\pi} \frac{\text{Im}\left[(Y_{\nu }^*)_{\alpha 2} (Y_{\nu}^{})_{\alpha 3} (Y_{\nu}^\dagger Y_{\nu}^{})_{23} \right]}{(Y_{\nu}^\dagger Y_{\nu}^{})_{22}} f(x_{32}^{}) \nonumber \\
& & +\frac{1}{8\pi} \frac{\text{Im}\left[(Y_{\nu}^*)_{\alpha 2} (Y_{\nu}^{})_{\alpha 3} (Y_{\nu}^\dagger Y_{\nu}^{})_{32} \right]}{(Y_{\nu}^\dagger Y_{\nu}^{})_{22}} g(x_{32}^{}),
\end{eqnarray}
where $x_{32}^{}=M_3^2/M_2^2$, and $f(x_{32}^{})$ and $g(x_{32}^{})$ are given by
\begin{eqnarray}
f(x_{32}^{}) &=& \sqrt{x_{32}^{}}\left[ (1+x_{32}^{})\ln\frac{1+x_{32}^{}}{x_{32}^{}} +\frac{1}{x_{32}^{}-1}-1\right], \\
g(x_{32}^{}) &=&  \frac{1}{x_{32}^{}-1}.
\end{eqnarray}
Note that we have neglected the contributions from $N_1^{}$, as $M_1^{} \ll M_2^{}$. Since leptogenesis of $N_2^{}$ operates in the two-flavor region, one then has $\epsilon_{2e +\mu} = \epsilon_{2 e}+\epsilon_{2\mu}$. 

On the other hand, the washout factor $K_{i\alpha}$ for an individual flavor $\alpha$ is given by

\begin{eqnarray}
K_{i\alpha} &=& K_i  P_{i\alpha},
\end{eqnarray}
where $P_{i\alpha}$ is the branching ratio for $N_i$ decaying into $l_\alpha$, namely, 
\begin{eqnarray}
P_{i\alpha}=\frac{|(Y_{\nu}^{})_{\alpha i}|^2}{\sum_\gamma |(Y_{\nu})_{\gamma i}|^2},
\end{eqnarray}
and $K_i^{}$ is the total washout factor for $N_i$. Such a total washout factor $K_i^{}$ can be evaluated according to
\begin{eqnarray} \label{eq:ki}
K_i \equiv \frac{\Gamma_i}{H(M_i)} = \frac{\tilde{m}_i}{\tilde{m}^*},
\end{eqnarray}
where $H(M_i)$ is the Hubble expansion rate at the temperature $T=M_i$, and two commonly used parameters $\tilde{m}_i$ and $\tilde{m}^*$ are introduced, with their definitions given by

\begin{eqnarray}\label{eq:mi}
\tilde{m}^* & \equiv & 8\pi \frac{v^2}{M_i^2}H(M_i) \simeq  10^{-3}~ \text{eV}, \\
\tilde{m}_i &\equiv & \frac{8\pi v^2}{M_i^2}\Gamma_i=\frac{(Y_{\nu}^\dagger Y_{\nu}^{})_{ii}^{} v^2}{M_i}.
\end{eqnarray}
Again, for $N_2^{}$ in the two-flavor region, we have $K_{2e+\mu}= K_{2e}+K_{2\mu}$.

With washout factors obtained, one then proceeds to calculate the efficiency factor $\kappa$, according to the following approximated formulae \cite{Blanchet:2006be}

\begin{eqnarray} \label{eq:kappa}
\kappa_{i\alpha} (K_{i\alpha}) \approx \frac{2}{K_{i\alpha}z_B^{}(K_{i\alpha})}\left[1-\text{exp}\left(-\frac{K_{i\alpha}z_B^{}(K_{i\alpha})}{2}\right)\right],
\end{eqnarray}
where $z_B^{}(x)=2+4~x^{0.13}\text{exp}(-2.5/x)$. 

So far we have presented the relevant formalism to calculate the asymmetry generated by $N_2^{}$ at $T \sim M_2^{}$. Such an asymmetry is then subject to the subsequent washout process led by $N_1^{}$. Since $M_1^{} \ll 10^9~\text{GeV}$, this $N_1^{}$-induced washout process would operate in the three-flavor region. We therefore first project $N_{\text{B-L}}^{T\sim M_2}$ into three flavors, and then let $N_1^{}$ to wash out those three individual flavor asymmetries according to its washout factors $K_{1\alpha}$. The final $(B-L)$ asymmetry is then found to be \cite{DiBari:2013qja}

\begin{eqnarray}\label{eq:N1washout}
N_{\text{B-L}}^{f} &=& \frac{K_{2e}}{K_{2e+\mu}}~\epsilon_{2e+\mu}~\kappa(K_{2e+\mu})~e^{-\frac{3\pi}{8}K_{1e}}+\frac{K_{2\mu}}{K_{2e+\mu}}~\epsilon_{2e+\mu}~\kappa(K_{2e+\mu})~e^{-\frac{3\pi}{8}K_{1\mu}} \nonumber \\
& & +\epsilon_{2\tau}~\kappa(K_{2\tau})~e^{-\frac{3\pi}{8}K_{1\tau}},
\end{eqnarray}
where the two ratios $K_{2e}^{}/K_{2e+\mu}$ and $K_{2\mu}/K_{2e+\mu}$ reflect the projection processes, and the suppression factor $e^{-3\pi K_{1\alpha}/8}$ is due to the washout process induced by $N_1^{}$.

Finally, the baryon-to-photon number ratio $\eta_B^{}$ is given by

\begin{eqnarray}
\eta_B^{} \approx 0.96\times 10^{-2} N_\text{B-L}^f,
\end{eqnarray}
and a successful leptogenesis is claimed when such a predicted value agrees with the actual observation from Planck \cite{Ade:2013zuv},

\begin{eqnarray}\label{eq:etab_ob}
\eta_B^{0}=(6.065 \pm 0.090)\times 10^{-10}.
\end{eqnarray}

\subsection{Qualitative Analysis}

We now begin to apply the above formalism into our previous set-up, namely, hierarchical neutrino Yukawa matrix $Y_\nu^{}$ with four texture zeros. A rough estimation on the final baryon asymmetry will be carried out first, followed by a careful examination of textures so as to identify those which are unlikely to reproduce the observed baryon asymmetry. 

Let us start with the asymmetry generated by $N_2^{}$ around $T \sim M_2$. With the hierarchical forms of $Y_\nu^{}$ and $M_R^{}$ given in Eq.(\ref{eq:YMR}), the asymmetry factor $\epsilon_{2\alpha}$ is approximately given by

\begin{eqnarray}
\epsilon_{2\alpha} \approx \frac{3}{16\pi}\frac{\text{Im}\left[(Y^*)_{\alpha 2} (Y^{})_{\alpha 3} (Y^\dagger Y^{})_{23} \right]}{(Y^\dagger Y)_{22}^{}(Y^\dagger Y)_{33}^{}}\lambda^{2x},
\end{eqnarray}
where we have used the approximations that $f(x_{32}^{}) \approx -3/(2\sqrt{x_{32}^{}})$ when $x_{32} \gg 1$, and that the term with $g(x_{32}^{})$ is neglected, as it is much smaller than the term with $f(x_{32}^{})$. Notice that for all textures, the asymmetry factor $\epsilon_{2\alpha}^{}$ is only non-zero along a single flavor. This is due to the fact that in order to have two or more non-zero $\epsilon_{2\alpha}$'s, one has to require the number of non-zero elements in the last two columns of $Y_\nu^{}$ to be five or more. The first column of $Y_\nu^{}$ would then be full of zeros, leading to a case that is equivalent to the minimal seesaw scenario with only two sets of RH neutrinos included. Because of such an observation, the non-zero $\epsilon_{2\alpha}^{}$ coincides with the total asymmetry factor $\epsilon_2^{}$ for $N_2^{}$, namely,

\begin{eqnarray}
\epsilon_{2\alpha} = \epsilon_2^{} \approx \frac{3}{16\pi}\frac{\text{Im}\left[(Y^\dagger Y^{})_{23}^2 \right]}{(Y^\dagger Y)_{22}^{}(Y^\dagger Y)_{33}^{}}\lambda^{2x},
\end{eqnarray}
and such a flavor index $\alpha$ can be identified by the fact that both $Y_{\alpha2}^{}$ and $Y_{\alpha 3}^{}$ entries need to be non-zero. 

Assuming that the non-zero elements of $Y$ are all around the same order, and that no accidental cancellation of phases exists, the factor $\text{Im}\left[(Y^\dagger Y^{})_{23}^2 \right]/[(Y^\dagger Y)_{22}^{}(Y^\dagger Y)_{33}^{}]$ is then of order $\lesssim \mathcal{O}(1)$.  The asymmetry factor $\epsilon_{2\alpha}$ is then mostly determined by the hierarchical factor $\lambda^{2x}$, which originates from our GUT requirements on $Y_\nu^{}$. With $x \sim 3.3$, we then have $\epsilon_{2\alpha} \lesssim \mathcal{O}(10^{-6})$.\footnote{Note that $y$ does not play a significant role in this $N_2^{}$-dominated leptogenesis. We may therefore relax the assumption on $\beta$ in the first place.}

As for the washout factors, one similarly plugs $Y_\nu^{}$ and $M_R^{}$ from Eq.(\ref{eq:YMR}) into Eq.(\ref{eq:ki}), yielding 

\begin{eqnarray}
K_i \approx \frac{(Y^\dagger Y)_{ii}^{}}{10^{-3}~\text{eV}}.
\end{eqnarray}
Again, since $M_\nu^{} = YY^T$, the size of $(Y^\dagger Y)_{ii}^{}$ is around the mass scale of light neutrinos. Given the large allowed range of the lightest neutrino mass identified in Fig.\ref{fg:LE}, $K_i$ can be from $\sim 1$ to $\sim 100$. Taking a typical value of $K_{i} =10$, the efficiency factor $\kappa$ is found to be around 0.03, according to Eq.(\ref{eq:kappa}). 

Assembling the above pieces together, the generated (B-L) asymmetry $N_\text{B-L}^{T\sim M_2}$ around the mass scale of $N_2^{}$ can be $\mathcal{O}(10^{-8})$. Were such an asymmetry not significantly washed out by $N_1^{}$, the finally generated baryon-to-photon number ratio $\eta_B^{}$ would be around $\mathcal{O}(10^{-10})$, the correct order to reproduce the observed value of $\eta_B^{}$. Therefore, the pressing question now is whether the asymmetry generated by $N_2^{}$ can survive the washout led by $N_1^{}$.

Fulfilling such a requirement actually can eliminate a number of textures of $Y_\nu^{}$. From Eq.(\ref{eq:N1washout}), one can see that as long as the washout factor $K_{1\alpha}$ is non-vanishing and of order $\mathcal{O}(10)$, the asymmetry generated by $N_2^{}$ along that flavor $\alpha$ would be sufficiently washed out by $N_1^{}$, due to the suppression factor $e^{-3\pi K_{1\alpha}/8}$. Therefore, for some textures of $Y_\nu^{}$, if $\epsilon_{2\alpha}^{} \neq 0$ and $K_{1\alpha}\neq 0$ hold simultaneously, they are unlikely to yield enough baryon asymmetry. Simply based on such a qualitative argument, without resorting to a detailed calculation, one is able to eliminate a number of textures of $Y_\nu^{}$, which are categorized into the following four classes. 

\begin{itemize}

\item $\mathbf{\epsilon_{2\tau} \neq 0 ~ (\epsilon_{2e}=0, \epsilon_{2\mu}=0)}$: One only has the asymmetry generated along the $\tau$ flavor. Hence as long as $(Y_\nu)_{\tau 1}$ is non-zero, which leads to a non-zero $K_{1\tau}$, very little asymmetry can survive eventually. In total we have six textures of this kind, and one of them is given by

\begin{eqnarray}
\begin{pmatrix}
0 & \times & 0 \\
0 & 0 & \times \\
\textbf{X} & \textbf{X} & \textbf{X}
\end{pmatrix},
\end{eqnarray}
where the crucial elements are highlighted in boldface. The other five can be obtained by distributing the two non-zero elements at different places in the first two rows, barring the cases that they are at the same column. 

\item $\mathbf{\epsilon_{2e} \neq 0~ (\epsilon_{2\mu}=0, \epsilon_{2\tau}=0)}$: This case is slightly different from the above one, and the number of unpromising textures is four. The shortage of two textures is due to the complexity introduced by the two-flavor approximation. Recall that when the asymmetry is first generated by $N_2^{}$, we are in the two-flavor approximation, where $e$ and $\mu$ flavors are indistinguishable. Therefore, even though $\epsilon_{2\mu}^{} =0$, as long as $K_{2\mu}$ is non-zero, we still have some projected asymmetry along the $\mu$ flavor. Then if $K_{1\mu}=0$ (equivalently $(Y_\nu)_{\mu 1}=0$) also holds, such a projected asymmetry along the $\mu$ flavor can survive, and may be enough to reproduce the observed baryon asymmetry. As a result, among those six would-be-eliminated textures, which can be obtained by exchanging the first and third rows of $Y_\nu^{}$ in the above case, two of them are actually promising ones, and they are given by

\begin{eqnarray}
\begin{pmatrix}
\textbf{X} & \textbf{X} & \textbf{X}\\
\textbf{0} & \textbf{X} & 0 \\
0 & 0 & \times 
\end{pmatrix}, \qquad 
\begin{pmatrix}
\textbf{X} & \textbf{X} & \textbf{X}\\
\textbf{0} & \textbf{X} & 0 \\
\times & 0 & 0 
\end{pmatrix}.
\end{eqnarray}

\item $\mathbf{\epsilon_{2\mu} \neq 0~ (\epsilon_{2e}=0, \epsilon_{2\tau}=0)}$: This case is similar to the above $\epsilon_{2e}^{} \neq 0$ case, with the first two rows of $Y_\nu^{}$ interchanged. 

\item $\mathbf{\epsilon_{2\alpha}=0}$: In this case, no asymmetry can be generated. It arises when the last two columns of $Y_\nu^{}$ are orthogonal, leading to a vanishing $(Y_\nu^\dagger Y_\nu^{})_{23}$. In total we find 18 textures of this kind, and they fall into two categories: 

\begin{itemize}
\item One and two non-zero elements are presented in the last two columns of $Y_\nu^{}$. There are 12 patterns in total. One of them can be

\begin{eqnarray}
\begin{pmatrix}
\times & \textbf{X} & \textbf{0}\\
\times & \textbf{X} & \textbf{0} \\
0 & \textbf{0} & \textbf{X} 
\end{pmatrix},
\end{eqnarray}
and the other eleven patterns can be obtained by exchanging the last two columns, and by shuffling elements in a particular columns, with the caution that some of them may lead to the cases that have three zeros in a row, or that have block diagonal forms, both of which are already out of our consideration in the first place. 

\item Both of the last two columns of $Y_\nu^{}$ contain one non-zero element. One example texture of this kind can be 

\begin{eqnarray}
\begin{pmatrix}
\times & \textbf{X} & \textbf{0}\\
\times & \textbf{0} & \textbf{X} \\
\times & \textbf{0} & \textbf{0} 
\end{pmatrix},
\end{eqnarray}
and there are another 5 textures, which can be obtained by shuffling elements in a particular column, with the same caution as that stated above. 

\end{itemize}

\end{itemize}

To summarize, simply based on a qualitative consideration of this $N_2^{}$-dominated leptogenesis, we have found 32 textures that are unlikely to reproduce the observed baryon asymmetry. Our classification, however, is centred on leptogenesis, different from our previous categorization of textures, which is based on the resulting behaviours on the effective neutrino mass matrix $M_\nu^{}$. Such two probes of textures are independent, and complementary to each other. For this reason, in Table \ref{tb:summary} we show the number of promising textures after applying the above leptogenesis requirement within the previous categorization. As one can see, applying leptogenesis, even at this qualitative level, has already suggested that about half of the textures that are allowed by low energy neutrino data are unlikely to achieve a successful leptogenesis. Such effectiveness of eliminating textures highlights the important role played by the flavor effects, and this work can be served as an explicit example of using flavored leptogenesis to restrict the \emph{form} of neutrino Yukawa matrix, not simply the parameter space of some free parameters. 

\begin{table}
\centering
\begin{tabular}{c | c  c c  | c  c c }
\hline
\hline
\multirow{2}{*}{} & \multicolumn{3}{c|}{NH} & \multicolumn{3}{c}{IH} \\
\cline{2-7}
& LE & LE+LG(QA) & LE+LG(NS) & LE & LE+LG(QA) & LE+LG(NS)  \\
\hline
\textbf{Class IA} & 0 & 0 & 0 & 18 & 10 & 0 \\
\textbf{Class IB} & 18 & 10 & 7 & 18 & 10 & 5 \\
\textbf{Class IC} & 18 & 8 & 4 & 18 & 8 & 5\\
\hline
\textbf{Class IIA} & 6 & 4 & 2 & 0 & 0 & 0\\
\textbf{Class IIB} & 6 & 4 & 1 & 6 & 4 & 1\\
\textbf{Class IIC} & 6 & 4 & 1 & 6 & 4 & 1\\
\hline
\hline
\end{tabular}
\caption{Summary of the number of allowed textures under various considerations. ``LE'' stands for the scenario with only low energy (LE) neutrino data considered, while a qualitative analysis (QA) and a numerical study (NS) of leptogenesis (LG) are also included in ``LE+LG(QA)" and "LE+LG(NS)", respectively.}
\label{tb:summary}
\end{table}

\subsection{Numerical Study}

In the previous subsection we have seen how such a $N_2^{}$-dominated leptogenesis can disfavor textures of $Y_\nu^{}$ at a qualitative level, we now perform a quantitative numerical study on that. Our numerical study aims at two goals: first, check the previous qualitative study, trying to see if indeed there is no parameter space for those unpromising textures; second, for those promising textures, identify the allowed parameter space for $m_{1,3}^{}$, $\delta$ and $M_{ee}^{}$. 

Since such a $N_2^{}$-dominated leptogenesis is sensitive to the textures of $Y_\nu^{}$, we have to study all the textures that are allowed by low energy neutrino data one by one. For a given texture, we first reconstruct its corresponding $M_\nu^{}$ by sampling those five measured neutrino parameters within their $3\sigma$ ranges, and choosing $m_{1,3}^{}$ and $\delta$ according to their allowed parameter space given in Fig.\ref{fg:LE}. With such a reconstructed $M_\nu^{}$, we solve for the rescaled Yukawa matrix $Y$. From $Y$, we further determine its corresponding $Y_\nu^{}$ and $M_R^{}$ according to Eq.(\ref{eq:YMR}). Finally, a leptogenesis evaluation with those hierarchical forms of $Y_\nu^{}$ and $M_R^{}$ is carried out, and we retain the input neutrino parameters and the form of working texture as a solution, if the finally obtained baryon-to-photon ratio satisfying $\eta_B^{} > 5\times 10^{-10}$. Here we have adopted a conservative requirement on $\eta_B^{}$, so as to account for some oversimplifications that may have been made in the previous leptogenesis formalism. 

Our numerical results are given as follows. First, in global we find 15 textures allowed for the case of NH, while 12 textures for IH. The numbers of survived textures in terms of different classes are given in Table \ref{tb:summary}, and their specific forms are listed in Appendix A. 

Second, our previous qualitative elimination of textures is almost confirmed, with the exception that for IH we find two would-be-eliminated textures that actually contain some parameter space to reproduce the observed baryon asymmetry. These two revived cases are IB9 and IC9, according to the classification given in Appendix A. Through a careful inspection, we find that such two textures are allowed because their washout processes led by $N_1^{}$ along the $\mu$ flavor can be very ineffective, as $K_{1\mu}$ can be as low as around 1. Such a small value of $K_{1\mu}$ is not thought to be generic, and therefore is omitted in our previous qualitative analysis.

Third, we find that among those promising textures, only about half of them have parameter space to generate enough baryon asymmetry. The allowed parameter space for $m_{1,3}^{}$, $\delta$ and $M_{ee}^{}$ for each allowed texture is given in Figs.2-5. Among these figures, Fig.\ref{fg:NHCIB} (Fig.\ref{fg:NHCIIAB}) is for Class IB (Classes IIA and IIB) in the case of NH, while Fig.\ref{fg:IHCIBIIB} gives the results for Classes IB and IIB in IH. Because $\theta_{23} \sim 45^\circ$, the allowed parameter space for textures in Class IC (IIC) can be roughly obtained by performing $\delta \rightarrow \delta + 180^\circ$ in their corresponding textures in Class IB (IIB), with the caution that for Class IC in NH, no corresponding textures of IB3, IB4 and IB7 are found. Furthermore, since in Fig.\ref{fg:IHCIBIIB} the allowed parameter space for Class IB in IH is so narrowly confined that we give a zoomed-in plot in Fig.\ref{fg:IHCIB}.

Similar to Fig.\ref{fg:LE}, we also only show the boundaries of allowed parameter space in these figures. Dashed contours represent the parameter space that is allowed by the low energy neutrino data for a given class, while solid contours are for individual textures within that given class. As one can see, the previously allowed parameter space by the low energy neutrino data is greatly shrunk after the consideration of leptogenesis. This highlights the constraining power of leptogenesis on parameter space, in addition to its eliminating power on textures.

Lastly, by combining Fig.\ref{fg:LE} with Table \ref{tb:summary}, one observes an interesting correlation between the lightest neutrino mass and the level of reduction of textures from QA to NS. It seems that among all cases the larger the lightest neutrino mass has to be, the greater reduction would exist. One extreme example is the case Class IA in IH, where although 10 promising textures can be identified in QA, none of them actually survives from NS. Such a sharp discrepancy is attributed to the fact that in this scenario the lightest neutrino mass has to be quite large, $m_3^{} \gtrsim 0.015~\text{eV}$ (see Fig.\ref{fg:LE}), causing $K_{2e+\mu} \gtrsim 50$ ($\kappa(K_{2e+\mu}) \lesssim 0.005$) for most of textures. Given the asymmetry factor being around $\mathcal{O}(10^{-6})$, a successful leptogenesis is then hard to achieve.  

We next discuss these figures, and focus on the testability of these allowed classes in upcoming neutrino experiments. On the one hand, in all allowed classes, we find that the lightest neutrino mass cannot be larger than $\sim 0.015~\text{eV}$, and that larger values of $M_{ee}^{}$ are not favored. Such two observations may be correlated, and both of them point to the fact that the scenario with three quasi-degenerate light neutrinos are not allowed by our current set-up.  Therefore, any positive indication on such a quasi-degenerate scenario from upcoming experiments would then disfavor our working assumptions. It is worthwhile to note that such testability is absent previously when considering a generic four texture zero scenario for $Y_\nu^{}$. While imposing additional GUT requirements and requiring a successful leptogenesis restrict the model, the testability is increased considerably. Occam's razor is at work and taking effect. 

On the other hand, it is also possible that the scenario with quasi-degenerate light neutrinos is not favored by upcoming experiments. The question then is whether one can still discriminate classes or even individual textures. For NH, due to the substantial overlap of parameter space among all allowed classes, telling them apart at the class level is already quite challenging, not to mention a distinguish of textures within a class. However, for the case of IH, the predictions on $M_{ee}^{}$ for Class IB (and IC) and Class IIB (and IIC) are disjoint, see Fig.\ref{fg:IHCIBIIB}. One may therefore distinguish them by a measurement of $M_{ee}^{}$. Moreover, the validity of Class IB (and IC) in IH can also be probed by a measurement of $\delta$, as its prediction on $\delta$ is narrowly centred around $90^\circ$ and $270^\circ$, a region that can be easily probed by upcoming long-baseline neutrino oscillation experiments.

\section{Summary and Conclusion}


In the spirit of Occam's razor, we have considered a restricted Type-I seesaw scenario with four texture zeros in the neutrino Yukawa matrix, in the \emph{weak basis} where both the charged-lepton Yukawa matrix and the Majorana mass matrix for right-handed neutrinos are diagonal and real. Although within such a weak basis,  four is the maximal number of texture zeros compatible with low energy neutrino data and the assumption that no neutrino masses vanishes, it still possesses two more free parameters, compared to the number of neutrino parameters that are well measured presently. As a result, the number of textures that are allowed by low energy neutrino data is huge (54 for NH and 66 for IH), and the parameter space for the lightest neutrino mass $m_{1,3}^{}$ and the Dirac CP-violating phase $\delta$ is so weakly constrained that an experimental test with upcoming neutrino experiments becomes quite challenging.

To further constrain the model, we impose additional GUT requirements on the neutrino Yukawa matrix, by demanding it to exhibit a similar hierarchical pattern to that in the up-type quark sector. Constraints from leptogenesis are then applied, and due to the hierarchy requirement, leptogenesis operates in a $N_2^{}$-dominated scenario, where the asymmetry is mostly generated by the next-to-lightest right-handed neutrino $N_2^{}$. Due to the requirement that the asymmetry generated by $N_2^{}$ should not be washed out subsequently by the lightest right-handed neutrino $N_1^{}$, such a $N_2^{}$-dominated leptogenesis is found to be very sensitive to the texture of neutrino Yukawa matrix. A qualitative analysis shows that about half of the textures that are allowed by low energy neutrino data are unpromising to reproduce the observed baryon asymmetry. A further numerical study confirms such an observation, and also indicates that among those promising ones, about another half of them are also not able to achieve a successful leptogenesis. In the end, we have 15 viable textures for NH and 12 for IH, and their allowed parameter space reveals that the scenario with three quasi-degenerate light neutrinos are disfavored. One may therefore exclude our model by a positive confirmation of such a quasi-degenerate scenario from upcoming neutrino experiments. 

Several avenues for future work could be as follows. First, one may relax the assumption that lighter eigenvalues in the up-type quark and neutrino Yukawa matrices are related by Georgi-Jarlskog type of relations. A  parameter scan on both $\alpha$ and $\beta$ can be done, and some general and more robust conclusions regarding such GUT requirements may be drawn. Second, as was done in \cite{Choubey:2008tb,Adhikary:2009kz}, one may further restrict the model by applying the constraint from charged lepton flavor violation, provided that a framework with supersymmetry is adopted. The number of allowed textures may get further reduced, as may be the allowed parameter space. Confirming or excluding it with upcoming neutrino experiments would become even more promising. Third, an inclusion of renormalization group running effects may be done, since we discuss leptogenesis at the mass scales of heavy RH neutrinos. Such a consideration is insignificant in our current SM set-up, due to the well-known fact that the RG evolution of neutrino parameters is very mild in the framework of the SM. However, if a supersymmetric framework were chosen, e.g., when considering the constraint from charged lepton flavor violation as was stated above, significant RG running effects might be present, especially in scenarios with large $\tan\beta$, or with quasi-degenerate light neutrinos in IH. A consistent study including RG effects would then be warranted. Finally, one may consider other minimal scenarios of Type-I seesaw mechanism but not working in the weak basis, i.e., $Y_e^{}$ and $M_R^{}$ are not restricted to be diagonal. A systematic study on allowed textures for all $Y_e^{}$, $Y_\nu^{}$ and $M_R^{}$ would then be worthwhile, in analogy to the similar work done previously in the quark sector \cite{Ramond:1993kv}.

\section*{Acknowledgement}

I am greatly thankful to Dr. Shun Zhou for useful discussions, making suggestions on the manuscript, and especially stimulating my interest in examining minimal models with the current neutrino data in the spirit of Occam's razor. I also wish to thank Michael Jay P\'erez and Gaoli Chen for commenting on the manuscript. Technical help from Yu Lu on using \texttt{Mathematica} is also highly appreciated. This work was supported in part by the Innovation Program of the Institute of High Energy Physics under Grant No. Y4515570U1.

\newpage
\appendix

\section{Allowed Textures}

In this appendix, we list all allowed textures within each allowed class for both NH and IH. Regardless of the mass hierarchy, all the allowed textures are given by

\begin{itemize}

\item \textbf{Class IB:}

\begin{eqnarray}
\textbf{IB1:}
\begin{pmatrix}
0 & \times & \times \\
\times & 0 & \times \\
\times & 0 & 0
\end{pmatrix},
\qquad 
\textbf{IB2:}
\begin{pmatrix}
\times & \times & 0 \\
0 & \times & \times \\
0 & 0 & \times
\end{pmatrix},
\qquad
\textbf{IB3:}
\begin{pmatrix}
0 & \times & 0 \\
\times & \times & \times \\
\times & 0 & 0
\end{pmatrix},
\nonumber \\
\textbf{IB4:}
\begin{pmatrix}
0 & \times & 0 \\
\times & \times & \times \\
0 & 0 & \times
\end{pmatrix},
\qquad 
\textbf{IB5:}
\begin{pmatrix}
\times & 0 & 0 \\
\times & 0 & \times \\
0 & \times & \times
\end{pmatrix},
\qquad
\textbf{IB6:}
\begin{pmatrix}
0 & \times & 0 \\
0 & \times & \times \\
\times & 0 & \times
\end{pmatrix},
\nonumber \\
\textbf{IB7:}
\begin{pmatrix}
0 & 0 & \times \\
0 & \times & \times \\
\times & \times & 0
\end{pmatrix},
\qquad 
\textbf{IB8:}
\begin{pmatrix}
0 & \times & \times \\
\times & \times & 0 \\
\times & 0 & 0
\end{pmatrix},
\qquad
\textbf{IB9:}
\begin{pmatrix}
\times & 0 & 0 \\
\times & \times & \times \\
0 & 0 & \times
\end{pmatrix}. \nonumber
\end{eqnarray}

\item \textbf{Class IC:} Taking out textures IB3, IB4 and IB7, and interchanging the last two rows of the remaining allowed textures in Class IB yield the six allowed textures in Class IC. Also, the numbering of textures in Class IC would follow that in Class IB, for example, texture IC8 indicates that it descends from IB8. 

\item \textbf{Class IIA:}

\begin{eqnarray}
\textbf{IIA1:}
\begin{pmatrix}
0 & 0 & \times \\
\times & 0 & \times \\
0 & \times & \times
\end{pmatrix},
\qquad
\textbf{IIA2:}
\begin{pmatrix}
0 & 0 & \times \\
0 & \times & \times \\
\times & 0 & \times
\end{pmatrix}. \nonumber
\end{eqnarray}

\item \textbf{Class IIB:}

\begin{eqnarray}
\textbf{IIB1:}
\begin{pmatrix}
\times & 0 & \times \\
0 & 0 & \times \\
0 & \times & \times
\end{pmatrix}. \nonumber
\end{eqnarray}

\item \textbf{Class IIC:} Similarly, the only allowed texture in Class IIC can be obtained by interchanging the last two rows of the texture in Class IIB.  

\end{itemize}

The allowed textures regarding a specific mass hierarchy are then given in Table \ref{tb:texture}. Numbers in the table refer to the above numbered textures. 

\begin{table}
\centering 
\begin{tabular}{c | c | c}
\hline 
\hline
& NH & IH \\
\hline
\textbf{Class IA} & - & -\\
\textbf{Class IB} & 1,2,3,4,5,6,7 & 1,2,5,8,9 \\
\textbf{Class IC} & 1,2,5,6 & 1,2,5,8,9 \\
\hline
\textbf{Class IIA} & 1,2 & - \\
\textbf{Class IIB} & 1 & 1 \\
\textbf{Class IIC} & 1 & 1\\
\hline
\hline
\end{tabular}
\caption{Allowed textures in all classes.}
\label{tb:texture}
\end{table}

\newpage

\begin{figure}
\centering
\includegraphics[scale=0.7]{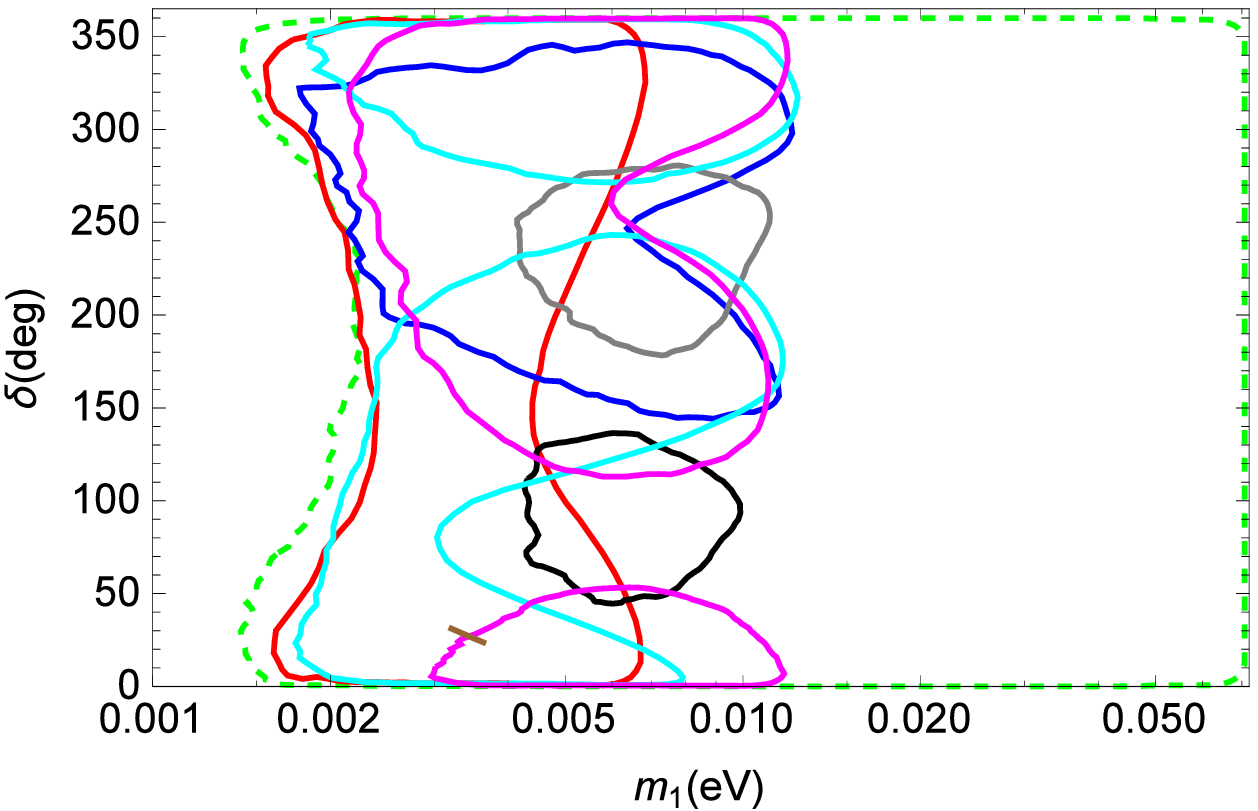}~~~~~~~~~~~~~~~~~~~~~~\\
\includegraphics[scale=1.0]{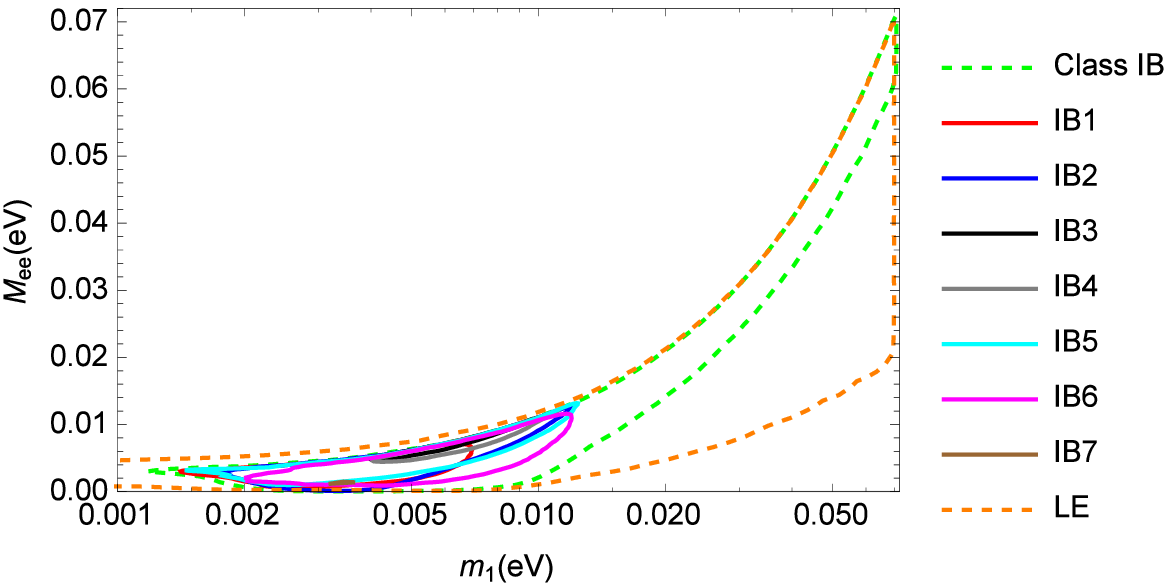}
\caption{(color online) Allowed parameter space for $m_1$, $\delta$ and $M_{ee}^{}$ in Class IB with NH. The orange dashed contour labelled by ``LE" is the same $3\sigma$ contour given in Fig.\ref{fg:LE}.}
\label{fg:NHCIB}
\end{figure}

\begin{figure}
\centering
\includegraphics[scale=0.7]{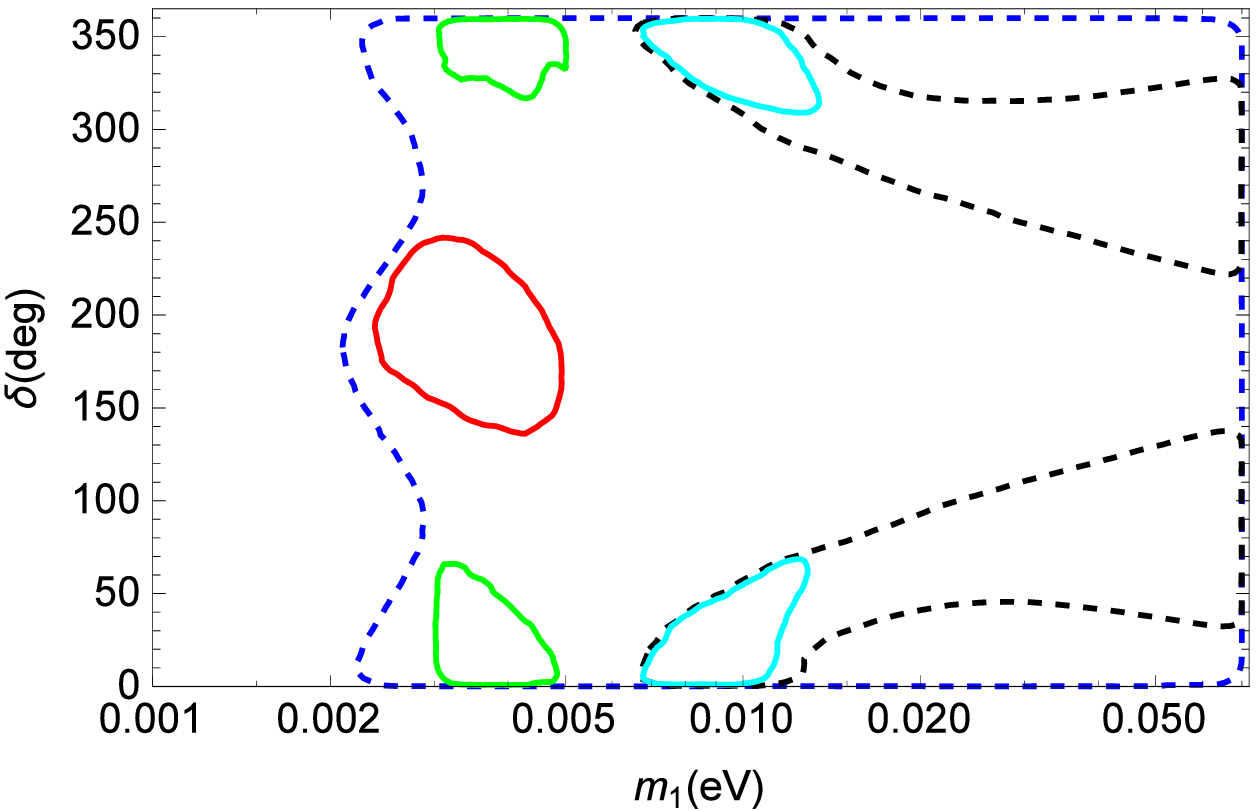}~~~~~~~~~~~~~~~~~~~~~~\\
\includegraphics[scale=1.0]{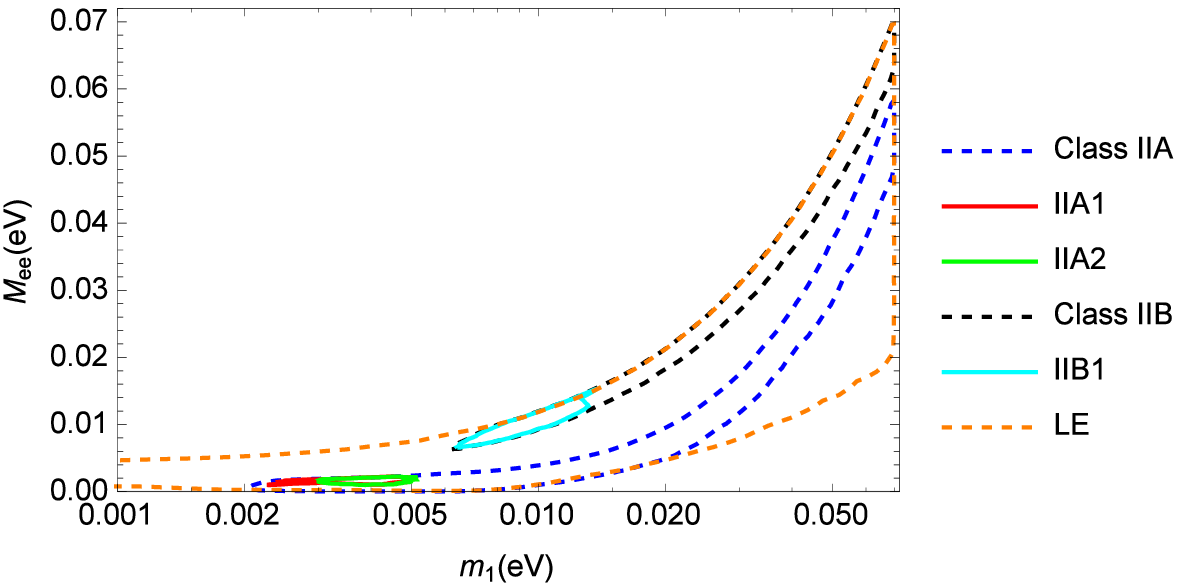}
\caption{(color online) Allowed parameter space for $m_1$, $\delta$ and $M_{ee}^{}$ in Classes IIA and IIB with NH.}
\label{fg:NHCIIAB}
\end{figure}

\begin{figure}
\centering
\includegraphics[scale=0.7]{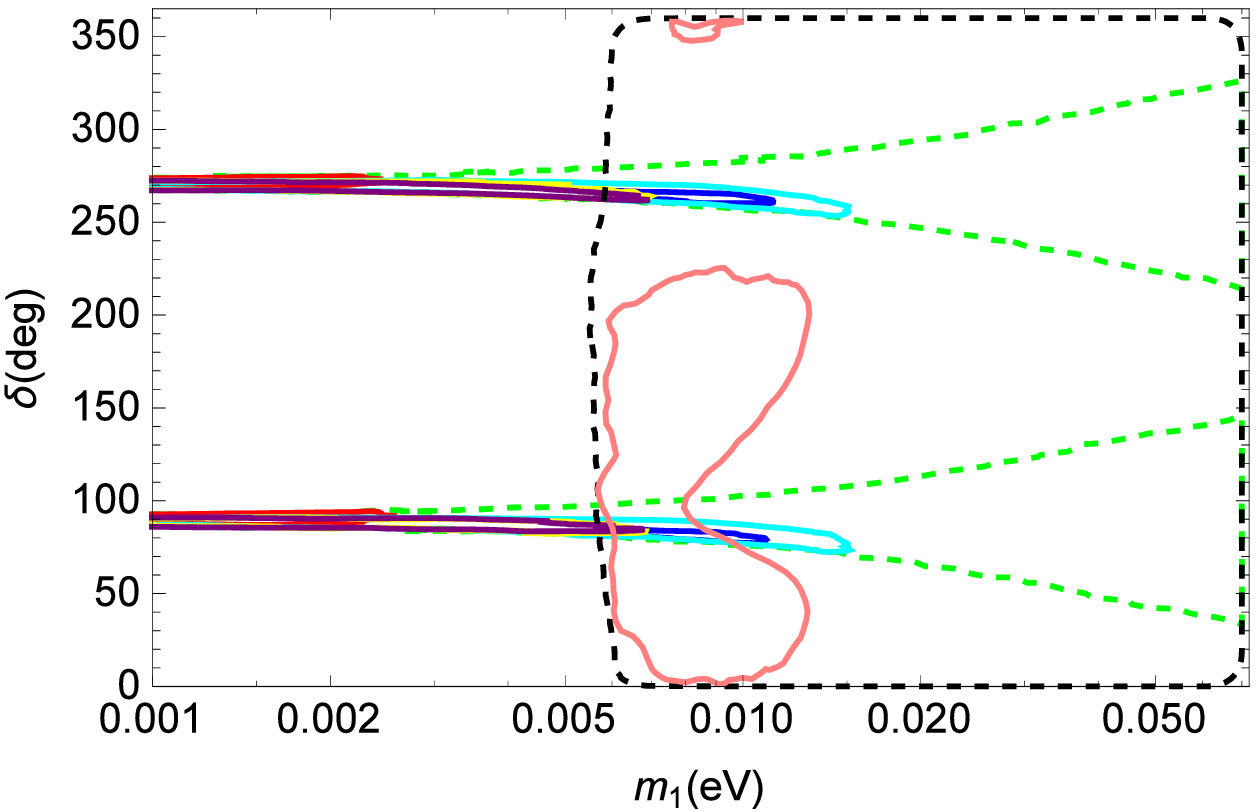}~~~~~~~~~~~~~~~~~~~~~~\\
\includegraphics[scale=1.0]{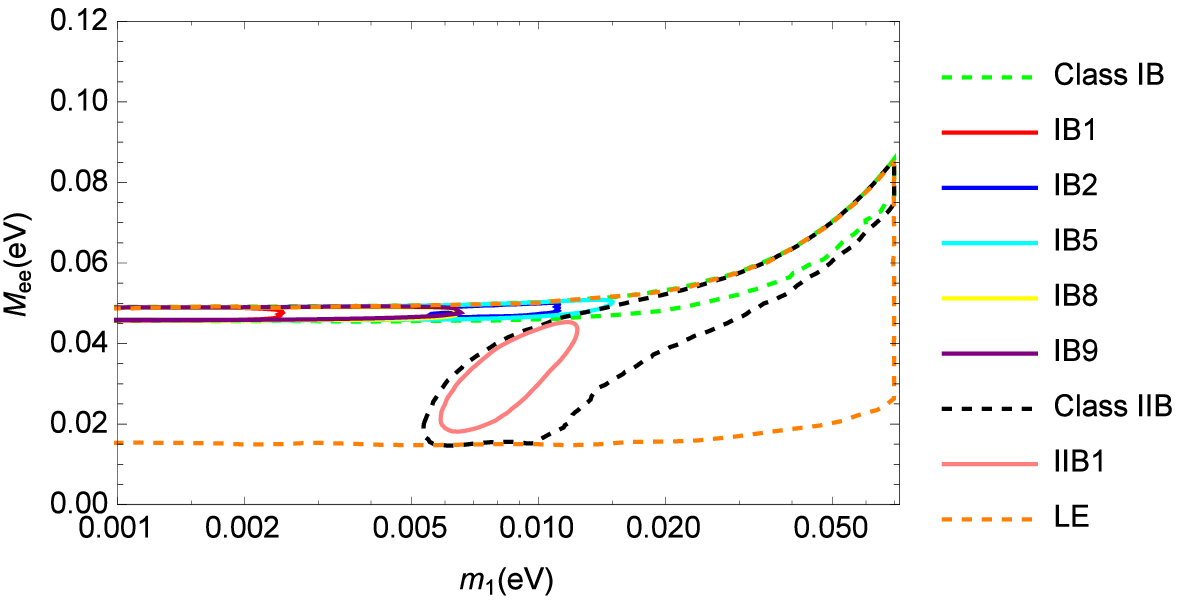}
\caption{(color online) Allowed parameter space for $m_3$, $\delta$ and $M_{ee}^{}$ in Classes IB and IIB with IH.}
\label{fg:IHCIBIIB}
\end{figure}

\begin{figure}
\centering
\includegraphics[scale=0.7]{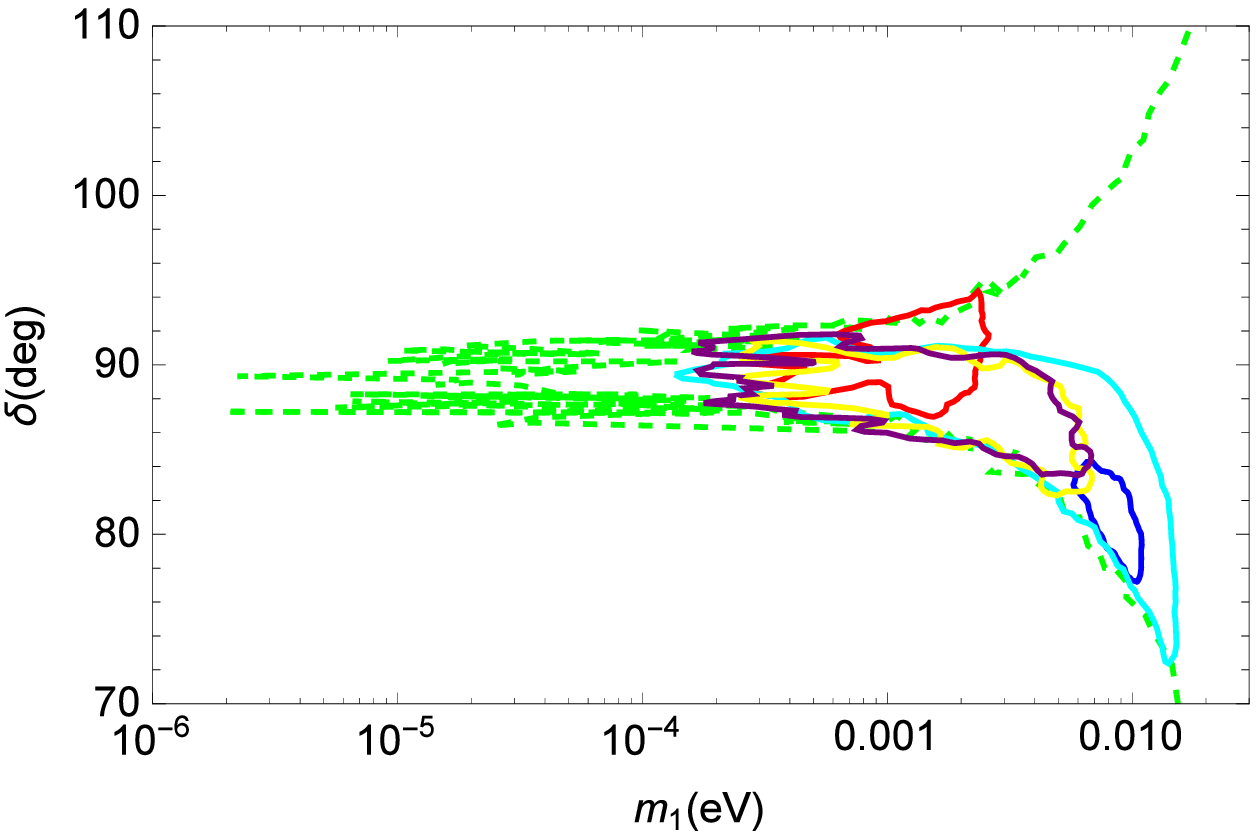}~~~~~~~~~~~~~~~~~~~~~~\\
\includegraphics[scale=1.0]{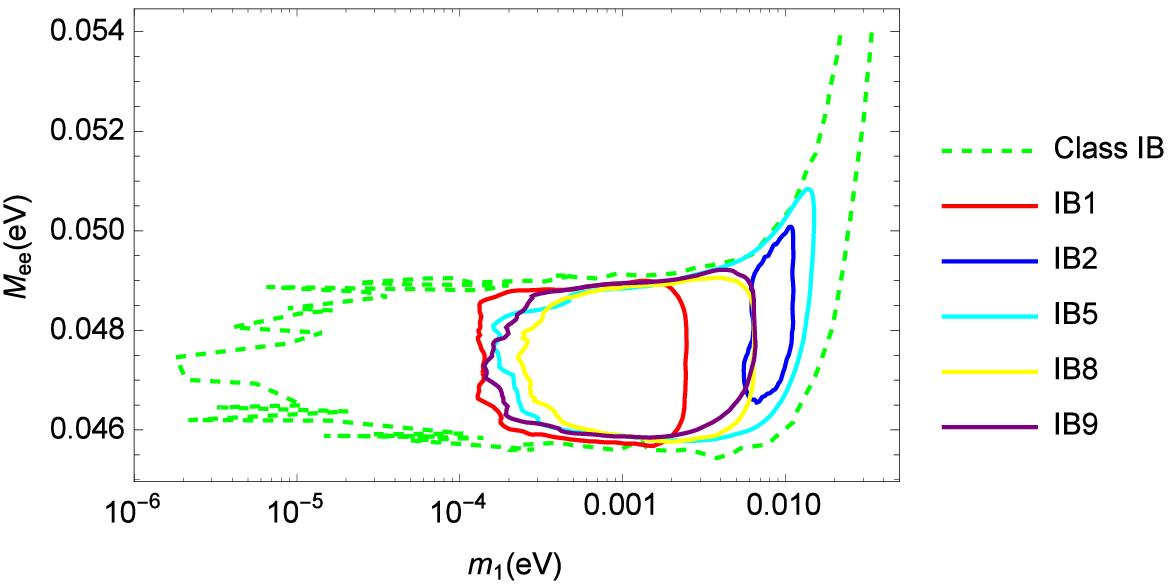}
\caption{(color online) Allowed parameter space for $m_3$, $\delta$ and $M_{ee}^{}$ in Class IB with IH. Note that there exists another branch of parameter space in the top panel with $\delta \rightarrow \delta + 180^\circ$.}
\label{fg:IHCIB}
\end{figure}

\end{document}